# Weak invariants of time-dependent quantum dissipative systems

Sumiyoshi Abe

*Department of Physical Engineering, Mie University, Mie 514-8507, Japan*
*and Institute of Physics, Kazan Federal University, Kazan 420008, Russia*

**Abstract.** The concept of weak invariant is introduced. Then, the weak invariants associated with time-dependent quantum dissipative systems are discussed in the context of master equations of the Lindblad type. In particular, with the help of the $su(1,1)$ Lie-algebraic structure, the weak invariant is explicitly constructed for the quantum damped harmonic oscillator with the time-dependent frequency and friction coefficient. This generalizes the Lewis-Riesenfeld invariant to the case of nonunitary dynamics in the Markovian approximation.

# I. INTRODUCTION

In most cases under realistic conditions, systems are not isolated but open, being surrounded by environments. Energy is exchanged between the objective and environmental systems, and physical coefficients characterizing the former one may vary explicitly in time. The dynamics governing such an open system is nonunitary in the quantum-mechanical regime.

In this work, we develop a discussion about the invariants of time-dependent quantum systems. Here, "time dependence" implies that physical coefficients contained in a system vary explicitly in time. In such a situation, we can define two different kinds of invariants: *strong and weak invariants*. Their definitions are presented. Then, we study the weak invariants of dissipative quantum open systems. To do so, we assume the dynamics of open systems to be Markovian, i.e., absence of long-term memories. This approximation drastically simplifies the discussion, since the master equation that is linear and preserves positive semidefiniteness of a density matrix necessarily becomes the Lindblad type [1,2]. Given Lindbladian operators that are responsible for dissipation, it is straightforward to derive the equation to be satisfied by the weak invariant operator, and therefore a main task is to solve such an operator equation. After general discussions about these issues, we explicitly construct for the time-dependent quantum damped harmonic oscillator the weak invariant together with the equation for an auxiliary *c*-number variable. This generalizes the celebrated Lewis-Riesenfeld invariant [3] of the time-dependent quantum harmonic oscillator without the damping term.

This paper is organized as follows. In Sec. II, the definitions of strong and weak invariants are presented. In Sec. III, a general discussion is made about the weak invariants associated with quantum master equations including, in particular, the Lindblad equation. In Sec. IV, which is the main part of the present work, the quantum damped harmonic oscillator with the time-dependent frequency and friction coefficient is analyzed, in detail. There the weak invariant, which generalizes the strong invariant of Lewis and Riesenfeld, is explicitly constructed with the help of the Lie-algebraic structure contained in the system. Section V is devoted to concluding remarks.

## II. DEFINITIONS OF STRONG AND WEAK INVARIANTS

Strong and weak invariants are defined as follows. A strong invariant is a Hermitian operator whose eigenvalues are all constant in time in terms of underlying quantum dynamics. On the other hand, a weak invariant, $\hat{I}(t)$, is a Hermitian operator whose eigenvalues are not constant in time, but its expectation value, $\langle \hat{I}(t) \rangle = \text{tr}\left[ \hat{I}(t)\, \hat{\rho}(t) \right]$, is conserved under time evolution of a system density matrix $\hat{\rho}(t)$ of an open system, the dynamics of which is nonunitary, in general.

For later convenience, let us recall a couple of examples of the strong invariants of the time-dependent quantum harmonic oscillator, the Hamiltonian of which reads

$$\hat{H}(t) = \frac{1}{2}\hat{p}^2 + \frac{1}{2}\omega^2(t)\hat{x}^2, \tag{1}$$

where $\hat{x}$ and $\hat{p}$ are the position and momentum operators in the Schrödinger picture, $\omega(t)$ the time-dependent frequency, and the mass is set equal to unity. The celebrated one of the quadratic form of the position and momentum operators is the Lewis-Riesenfeld invariant presented in Ref. [3]:

$$\hat{I}_0(t) = \frac{1}{2}\left[\left(\rho_0 \hat{p} - \dot{\rho}_0 \hat{x}\right)^2 + \frac{\hat{x}^2}{\rho_0^2}\right], \tag{2}$$

where $\rho_0 = \rho_0(t)$ is a $c$-number quantity obeying the auxiliary equation

$$\ddot{\rho}_0 + \omega^2(t)\rho_0 = \frac{1}{\rho_0^3} \tag{3}$$

with the overdots standing for time derivatives. Then, $\hat{I}_0(t)$ satisfies the following operator equation:

$$i\frac{\partial \hat{I}_0(t)}{\partial t} - \left[\hat{H}(t), \hat{I}_0(t)\right] = 0, \tag{4}$$

provided that here and hereafter $\hbar$ is set equal to unity. From Eq. (4), we can find that the eigenvalues of $\hat{I}_0(t)$ are constant in time (see the next section). The other example of the strong invariant [4], which is linear in the position and momentum, is $\hat{A}(t) = \varepsilon \hat{p} - \dot{\varepsilon}\hat{x}$, where $\varepsilon = \varepsilon(t)$ is a $c$-number quantity that satisfies the equation of the same form as Eq. (4), i.e., $\partial \hat{A}(t)/\partial t - [\hat{H}(t), \hat{A}(t)] = 0$, if $\varepsilon$ is a solution of the auxiliary equation, $\ddot{\varepsilon} + \omega^2(t)\varepsilon = 0$. Then, the eigenvalues of $\hat{A}(t)$ are also constant in time. This invariant is analogous to the Wronskian.

### III. WEAK INVARIANT OF TIME-DEPENDENT QUANTUM DISSIPATIVE SYSTEM

Consider a *generic* quantum open system whose Hamiltonian depends explicitly on time, $\hat{H} = \hat{H}(t)$. Since a system of our interest is dissipative, the dynamics governing it is nonunitary, and therefore $\hat{H}(t)$ does not fully describe time evolution of a quantum state of the system, in general. (It is known that there are Hamiltonian approaches to dissipative systems [5-7], but our viewpoint here is different from theirs.) Let $\hat{\rho}$ be a density matrix of a state of the system that is a positive semidefinite operator satisfying the normalization condition (i.e., $\text{tr}\,\hat{\rho} = 1$). Its time evolution may be written as follows:

$$i\frac{\partial \hat{\rho}}{\partial t} = \pounds(\hat{\rho}), \tag{5}$$

where $\pounds$ is a certain linear superoperator and may contain the commutator with $\hat{H}(t)$. A quantity, $\hat{I}(t)$, is a weak invariant associated with Eq. (5), if it obeys

$$i\frac{\partial \hat{I}(t)}{\partial t} + \pounds^*(\hat{I}(t)) = 0, \tag{6}$$

where £* is the adjoint of £. If $\hat{I}(t)$ satisfies Eq. (6), then its expectation value is, in fact, constant in time.

Let us assume that the dynamics governing the system is Markovian and preserves the positive semidefiniteness of as well as the normalization condition on $\hat{\rho}$. Then, Eq. (5) necessarily takes the Lindblad form [1,2]:

$$i\frac{\partial \hat{\rho}}{\partial t} = \left[\hat{H}(t), \hat{\rho}\right] - i\sum_n \alpha_n \left(\hat{L}_n^\dagger \hat{L}_n \hat{\rho} + \hat{\rho} \hat{L}_n^\dagger \hat{L}_n - 2\hat{L}_n \hat{\rho} \hat{L}_n^\dagger\right). \tag{7}$$

Here, $\alpha_n$'s are *non-negative* c-number coefficients and $\hat{L}_n$'s are referred to as the Lindbladian operators that may also depend on time. The Hamiltonian generates the unitary part of time evolution, whereas the dissipative nature of the system is described by the second term on the right-hand side. Then, Eq. (6) becomes

$$i\frac{\partial \hat{I}(t)}{\partial t} - \left[\hat{H}(t), \hat{I}(t)\right] - i\sum_n \alpha_n \left(\hat{L}_n^\dagger \hat{L}_n \hat{I}(t) + \hat{I}(t)\hat{L}_n^\dagger \hat{L}_n - 2\hat{L}_n^\dagger \hat{I}(t)\hat{L}_n\right) = 0, \tag{8}$$

which turned out to have been presented in Ref. [8].

In contrast to the unitary case, the eigenvalues of $\hat{I}(t)$ generically depend on time. To see it, let us consider the instantaneous orthonormal eigenstates $\{|\lambda_n, t\rangle\}_n$ satisfying

$$\hat{I}(t)|\lambda_n, t\rangle = \lambda_n(t)|\lambda_n, t\rangle. \tag{9}$$

Then, from Eq. (8), it follows that

$$\frac{d\lambda_i(t)}{dt} = 2 \sum_n \alpha_n \left( \lambda_i(t) \langle \lambda_i, t | \hat{L}_n^\dagger \hat{L}_n | \lambda_i, t \rangle - \langle \lambda_i, t | \hat{L}_n^\dagger \hat{I}(t) \hat{L}_n | \lambda_i, t \rangle \right), \quad (10)$$

which does not vanish, in general. Clearly, if the third-term on the left-hand side in Eq. (8) is absent, then the eigenvalues are constant in time. $\hat{I}(t)$ in such a case becomes reduced to the strong invariant.

## IV. WEAK INVARIANT OF QUANTUM DAMPED HARMONIC OSCILLATOR

Let us discuss the weak invariant of the quantum damped harmonic oscillator with the time-dependent frequency and friction coefficient. The Hamiltonian to be considered is the one in Eq. (1). Here, we rewrite it as follows:

$$\hat{H}(t) = \hat{K}_1 + \omega^2(t) \hat{K}_2. \quad (11)$$

Here, $\hat{K}_1$ and $\hat{K}_2$ together with $\hat{K}_3$ are the operators defined by

$$\hat{K}_1 = \frac{1}{2} \hat{p}^2, \qquad \hat{K}_2 = \frac{1}{2} \hat{x}^2, \qquad \hat{K}_3 = \frac{1}{2} \left( \hat{p} \hat{x} + \hat{x} \hat{p} \right). \quad (12)$$

These operators satisfy the following set of commutation relations:

$$\left[\hat{K}_1, \hat{K}_2\right] = -i\hat{K}_3, \qquad \left[\hat{K}_2, \hat{K}_3\right] = 2i\hat{K}_2, \qquad \left[\hat{K}_3, \hat{K}_1\right] = 2i\hat{K}_1, \qquad (13)$$

which is formally isomorphic to the *su*(1, 1) Lie algebra.

Let $\hat{Q}$ be a certain observable in the Schrödinger picture. From Eq. (7), time evolution of its expectation value is found to be given by

$$\frac{d\langle\hat{Q}\rangle}{dt} = i\langle[\hat{H}(t), \hat{Q}]\rangle - \sum_n \alpha_n \langle(\hat{L}_n^\dagger \hat{L}_n \hat{Q} + \hat{Q}\hat{L}_n^\dagger \hat{L}_n - 2\hat{L}_n \hat{Q}\hat{L}_n^\dagger)\rangle. \qquad (14)$$

Now, from Eq. (8) and the Lie-algebraic structure in the Hamiltonian in Eq. (11), it turns out to be sufficient to employ the following single Lindbladian operator:

$$\hat{L} \equiv \hat{L}_1 = a_1(t)\hat{K}_1 + a_2(t)\hat{K}_2 + a_3(t)\hat{K}_3, \qquad (15)$$

where $a(t)$'s are real *c*-number functions of *t*, and therefore $\hat{L}$ is a time-dependent Hermitian operator. Thus, the Lindblad equation in Eq. (7) is simplified to be

$$i\frac{\partial \hat{\rho}}{\partial t} = [\hat{H}(t), \hat{\rho}] - i\alpha(t)[\hat{L}, [\hat{L}, \hat{\rho}]], \qquad (16)$$

where $\alpha(t) \equiv \alpha_1(t) \geq 0$. Accordingly, Eq. (8) for a weak invariant becomes

$$i\frac{\partial \hat{I}(t)}{\partial t} - [\hat{H}(t), \hat{I}(t)] - i\alpha(t)[\hat{L}, [\hat{L}, \hat{I}(t)]] = 0. \qquad (17)$$

From the structure of the second term on the right-hand side of Eq. (16), it is clear that one of $a(t)$'s can be absorbed into $\alpha(t)$. Here, we may choose $a_1(t)$ to be so and set it as

$$a_1(t) = 1, \qquad (18)$$

without losing generality.

Our idea is to realize the equation of motion of the damped harmonic oscillator for the expectation values:

$$\frac{d^2\langle\hat{x}\rangle}{dt^2} + 2\kappa(t)\frac{d\langle\hat{x}\rangle}{dt} + \Omega^2(t)\langle\hat{x}\rangle = 0, \qquad (19)$$

where $\kappa(t)$ and $\Omega(t)$ are the time-dependent friction coefficient and modulated frequency to be discussed later, respectively. For $\hat{Q}$ to be $\hat{x}$ and $\hat{p}$ in Eq. (14) with Eq. (15), we have

$$\frac{d\langle\hat{x}\rangle}{dt} = \langle\hat{p}\rangle - \kappa(t)\langle\hat{x}\rangle, \qquad (20)$$

$$\frac{d\langle\hat{p}\rangle}{dt} = -\omega^2(t)\langle\hat{x}\rangle - \kappa(t)\langle\hat{p}\rangle, \qquad (21)$$

where

$$\kappa(t) = \alpha(t)\left(a_2(t) - a_3^2(t)\right). \qquad (22)$$

Therefore, we obtain Eq. (19) with

$$\Omega^2(t) = \omega^2(t) + \kappa^2(t) + \dot{\kappa}(t). \tag{23}$$

One may require both $\kappa(t)$ and $\Omega^2(t)$ to be non-negative. Since $\alpha(t) \geq 0$, the condition

$$a_2(t) - a_3^2(t) \geq 0 \tag{24}$$

should be fulfilled, in order for $\kappa(t)$ to be non-negative.

Here, it is worth recalling the Lewis-Riesenfeld strong invariant in Eq. (2) that corresponds to the case $\alpha(t) = 0$. The crucial point to be noted is that the time-dependent physical coefficient, $\omega(t)$, does not explicitly appear in Eq. (2): it shows up only in the auxiliary equation (3). We maintain this feature in generalizing the Lewis-Riesenfeld invariant to the dissipative dynamics. That is, *the form of the invariant should be kept unchanged but the auxiliary equation has to be modified*.

Let us write the weak invariant of the quantum damped harmonic oscillator as follows:

$$\hat{I}(t) = \rho^2 \hat{K}_1 + \left( \dot{\rho}^2 + \frac{1}{\rho^2} \right) \hat{K}_2 - \rho \dot{\rho} \hat{K}_3, \tag{25}$$

where $\rho = \rho(t)$ is a *c*-number quantity (which should not be confused with the density matrix, $\hat{\rho}$). This is, in fact, of the same form as Eq. (2). This quantity has to satisfy Eq. (17). Substituting Eqs. (11), (15), and (25) into Eq. (17), and using the algebra in Eq.

(13), we obtain the following coupled equations:

$$\kappa(t)\rho^2 - \alpha(t)\left[a_3^2\rho^2 + 2a_3\rho\dot{\rho} + \left(\dot{\rho}^2 + \frac{1}{\rho^2}\right)\right] = 0, \qquad (26)$$

$$\dot{\rho}\left(\ddot{\rho} + \omega^2(t)\rho - \frac{1}{\rho^3}\right)$$

$$+ \left\{\alpha\left[a_2^2\rho^2 + 2a_2a_3\rho\dot{\rho} + a_3^2\left(\dot{\rho}^2 + \frac{1}{\rho^2}\right)\right] - \kappa(t)\left(\dot{\rho}^2 + \frac{1}{\rho^2}\right)\right\} = 0, \qquad (27)$$

$$\rho\left(\ddot{\rho} + \omega^2(t)\rho - \frac{1}{\rho^3}\right) - \alpha\left[a_2a_3\rho^2 + 2a_2\rho\dot{\rho} + a_3\left(\dot{\rho}^2 + \frac{1}{\rho^2}\right)\right] = 0. \qquad (28)$$

Combining these equations with Eq. (22), we find

$$\alpha = \frac{4\kappa(t)\rho^4}{(\rho\dot{\rho})^2 + 4}, \qquad (29)$$

$$a_2 = \frac{\dot{\rho}^2}{2\rho^2} + \frac{1}{\rho^4}, \qquad (30)$$

$$a_3 = -\frac{\dot{\rho}}{2\rho}, \qquad (31)$$

and the auxiliary equation

$$\ddot{\rho} - \kappa(t)\dot{\rho} + \omega^2(t)\rho = \frac{1}{\rho^3}. \qquad (32)$$

Therefore, the Lindbladian operator in Eq. (15) is fully determined, and the weak invariant is given in the form in Eq. (25) with the auxiliary equation being generalized from Eq. (3) to Eq. (32), now. This is the main result of the present work.

An intriguing point is that the signs of the friction terms in Eqs. (19) and (32) are opposite. This *time reversal* structure reminds one of the work in Ref. [9] (see also Ref. [7]), although the present approach is radically different from it.

Finally, let us consider the situation that both $\kappa(t)$ and $\omega(t)$ slowly vary in time. In this case, Eq. (32) has the solution of the following form:

$$\rho = \frac{1}{\omega^{1/2}(t)} - \frac{\kappa(t)}{8\omega^{7/2}(t)}\dot{\omega}(t) - \frac{1}{16\omega^{9/2}(t)}\left\{3 - \frac{7}{4}\left[\frac{\kappa(t)}{\omega(t)}\right]^2\right\}\dot{\omega}^2(t)$$

$$-\frac{\kappa(t)}{32\omega^{11/2}(t)}\dot{\kappa}(t)\dot{\omega}(t) + \frac{1}{8\omega^{7/2}(t)}\left\{1 - \frac{1}{4}\left[\frac{\kappa(t)}{\omega(t)}\right]^2\right\}\ddot{\omega}(t) + \cdots\cdots. \quad (33)$$

This expression systematically determines corrections to the adiabatic approximation.

## V. CONCLUDING REMARKS

We have defined the strong and weak invariants of time-dependent quantum systems. We have studied the weak invariants of time-dependent quantum dissipative systems based on the Lindblad equation. We have explicitly constructed the weak invariant for the quantum damped harmonic oscillator, whose frequency and friction coefficient depend on time. In this way, we have generalized the Lewis-Riesenfeld strong invariant to the case of the nonunitary dissipative dynamics. We have also observed that an intriguing structure exists in respect of the time reversal symmetry.

The method of invariants in time-dependent quantum systems *without dissipations* has found a variety of applications to constructions of the coherent and squeezed states [10-13], geometric phases [14-16], fermionic systems [17], charged quantum fields in time-dependent external electromagnetic fields [18], quantum fields in cosmological backgrounds [19,20], quantum computation [21], and quantum cosmology [22], to name but a few. It is also known in classical theory without dissipations [23] (see also Ref. [24] for a simplified explanation) that existence of such invariants can be understood from the viewpoint of Noether's theorem [25]. On the other hand, dissipative systems do not have Lagrangians or Hamiltonians, in general, unless spaces of dynamical variables are extended. In Ref. [26], it is shown to be possible to construct conserved quantities for classical non-Lagrangian—non-Hamiltonian systems (see also Refs. [27-30]). However, quantization of such systems is unclear, and therefore a connection between these discussions and the present work is yet to be clarified.

*Note added*.   Recently, we became aware of Ref. [31], in which an invariant different from the one presented here is discussed for the quantum damped harmonic oscillator.

**ACKNOWLEDGMENTS**

The author would like to thank Professor V. V. Dodonov for informing him about Ref. [31]. This work was supported in part by a Grant-in-Aid for Scientific Research from the Japan Society for the Promotion of Science (No. 26400391) and by the Program of Competitive Growth of Kazan Federal University from the Ministry of Education and Science of the Russian Federation.

———————————